# Tumor containment as an anti-percolation process

Arturo Tozzi (corresponding author)
ASL Napoli 1 Centro, Distretto 27, Naples, Italy
Via Comunale del Principe 13/a 80145
tozziarturo@libero.it

## ABSTRACT

Percolation theory from statistical physics has been applied to several aspects of tumor progression. Tumor growth on percolation clusters has been used to model spatial expansion, vascular percolation to describe nutrient supply and transport-related percolation to investigate drug and gene delivery. At the molecular level, mutational percolation has been employed to account for the emergence of malignant phenotypes, while inverse percolation to represent treatment-induced structural disruption. We examined whether tumor containment can be interpreted as an anti-percolation problem, in which spatial expansion depends on the formation of a connected malignant domain. We implemented a spatial simulation with biologically scaled parameters to represent tissue heterogeneity, local growth, cell movement and clearance. We measured both total malignant area and connectivity metrics, including the largest connected component and the probability of forming a spanning cluster. Our results indicate that tumor size and spatial connectivity are partially independent, with configurations of similar size showing different connectivity patterns. A transition from fragmented to connected structures emerged within a limited parameter range, consistent with a threshold-like behavior. Incorporating spatial connectivity into quantitative analysis, our approach provides a complementary way to characterize tumor organization. Potential applications include integration of structural descriptors into computational models of tumor growth, design of experimental systems to probe spatial organization and interpretation of therapeutic approaches via connectivity-based metrics.

## INTRODUCTION

Percolation frameworks have been applied to diverse aspects of tumor biology, including growth, transport, mutation and therapy response. Studies have shown that tumor hypoxia and radiotherapy response can be modeled through inverse percolation processes describing network fragmentation under treatment (Dimou et al. 2022), while tumor proliferation and invasion depend on the connectivity of permissive environments represented as percolation clusters (Jiang et al. 2016; Yang et al. 2022). Vascular architecture and perfusion have similarly been interpreted through invasion percolation models, linking irregular vessel networks to nutrient and drug delivery efficiency (Baish et al. 1996; Lee et al. 2006), while imaging studies have confirmed fractal and percolation-like features in tumor blood flow (Craciunescu et al. 1999). At smaller scales, cooperative mutational events have been described as forming percolating clusters associated with cancer hallmarks (Shin et al. 2017), while increasing cell density induces structural and dynamical percolation transitions in tumor cell assemblies (Liu et al. 2022). Percolation concepts also govern therapeutic penetration, as effective gene delivery and nanoparticle distribution depend on their ability to cross connected pathways within tumor tissues and lymphatic structures (Han et al. 2009; Xu et al. 2022). Still, extensions incorporating spatial gradients further highlight how microenvironmental factors can modulate local connectivity (Lambrou and Argyrakis 2026).

We suggest to assess tumor containment in terms of an anti-percolation problem (Notarmuzi et al. 2021; Mello et al. 2021; Brunk and Twarock 2021; Bianconi and Dorogovtsev 2024; Cirigliano et al. 2024), in which therapeutic success corresponds to maintaining large-scale connectivity and keeping malignant connectivity below a critical threshold. In a percolation framework, tumor expansion requires that local regions of proliferation, migration and matrix remodeling become connected enough to allow the emergence of a system-spanning structure. We suggest that the effective connectivity of this system could depend on multiple interacting factors, including cell density, extracellular matrix permissiveness, vascular support and immune-mediated removal. When these factors collectively exceed a critical level, isolated clusters merge into a coherent invasive domain; below this level, growth remains confined to disconnected regions. Therefore, the objective of therapy is shiftet toward maintaining the system in a subcritical regime, where malignant clusters cannot coalesce into a globally expanding structure. Interventions altering matrix organization, reducing motility or increasing local clearance will not necessarily erase tumor cells, but will fragment the spatial network of malignant elements, preventing the formation of a spanning cluster and stabilizing the system in a contained state of spatially fragmented domains.

We will proceed as follows. First, we formalize the connectivity-based description of tumor growth. Then, we develop computational models linking biological parameters to spatial organization. Finally, we present simulations and discuss experimental implications for containment strategies.

## METHODS

We studied whether spatial tumor containment can be represented as an anti-percolation process in which malignant expansion depends on the emergence of a connected cancer-cell domain. We used a two-dimensional stochastic lattice simulation, where each lattice element represented a square tissue patch of 20 μm per side, approximately corresponding to the diameter of one cell. We modeled tissue permissiveness, malignant colonization, motility-driven local expansion and clearance-mediated removal. We aimed to identify an experimentally discriminable signature, i.e., the separation between tumor size and tumor connectivity, such that two systems with comparable malignant area may differ in their ability to form a spanning invasive cluster.

**Spatial domain**. The simulated tissue was represented as a square two-dimensional domain $\Omega \subset \mathbb{R}^2$ with side length $L = 3.2$ mm in the main simulations. The domain was discretized into a regular lattice of $N \times N$ sites, with $N = 160$ and lattice spacing $\Delta x = 20\ \mu\text{m} = 0.02$ mm. Each site therefore represented an area $\Delta A = (\Delta x)^2 = 4 \times 10^{-4}$ mm$^2$. A lattice site was indexed by $i, j$, with spatial coordinates $x_i = i\Delta x$ and $y_j = j\Delta x$. At each time step $t$, the malignant state was defined by a binary variable $M_{ij}(t)$, where $M_{ij}(t) = 1$ denoted malignant occupation and $M_{ij}(t) = 0$ denoted absence of malignant occupation. The tissue-permissive state was represented by a second binary variable $P_{ij}$, where $P_{ij} = 1$ denoted a site permissive to tumor occupation and $P_{ij} = 0$ a non-permissive site. Tumor growth was therefore constrained by $M_{ij}(t) \leq P_{ij}$, so that malignant cells could expand only into locally admissible tissue positions.
To reduce dependence on the regular lattice assumption and approximate more realistic tissue organization, we introduced an alternative graph-based representation in which lattice sites were embedded into a disordered network with variable local connectivity. Starting from the same spatial coordinates, adjacency relations were modified by probabilistically removing or adding edges according to a distance-dependent rule, so that the neighborhood of each site was no longer restricted to a fixed von Neumann structure. Specifically, for any pair of sites $(i, j)$ and $(k, l)$, a connection was allowed with probability

$$\pi_{(ij),(kl)} = \exp\left( -\frac{d_{(ij),(kl)}}{\lambda}\right),$$

where $d_{(ij),(kl)}$ is the Euclidean distance and $\lambda$ is a characteristic interaction length. Malignant expansion and connectivity analysis were then repeated on this irregular graph, preserving the same growth and clearance dynamics but altering the topology of accessible paths. This procedure enabled comparison between regular and disordered spatial substrates, ensuring that the emergence of connected malignant domains was not restricted to a specific lattice geometry but depended on the broader structure of admissible connections.

**Permissive tissue**. Spatial heterogeneity was introduced by generating a smoothed random field $H_{ij}$ over the lattice. An initial field $R_{ij}$ was sampled from a uniform distribution, $R_{ij} \sim U(0,1)$. Spatial correlation was introduced by repeated nearest-neighbor averaging,

$$H_{ij}^{(k+1)} = \frac{1}{5}\left(H_{ij}^{(k)} + H_{i+1,j}^{(k)} + H_{i-1,j}^{(k)} + H_{i,j+1}^{(k)} + H_{i,j-1}^{(k)}\right),$$

with periodic array shifts used only to smooth the numerical field. After five smoothing iterations, the continuous field was thresholded to impose a specified permissive tissue fraction $p$. The threshold $\theta_p$ was defined as the empirical quantile satisfying

$$\Pr\left(H_{ij} \geq \theta_p\right) = p.$$

The permissive mask was then assigned as

$$P_{ij} = \begin{cases} 1, & H_{ij} \geq \theta_p, \\ 0, & H_{ij} < \theta_p. \end{cases}$$

In the parameter sweeps, $p$ was varied over values between 0.35 and 0.80, corresponding to 35 to 80 percent permissive tissue. This parameter was used as a compact representation of local tissue properties that enable malignant occupation, including matrix porosity, path availability and local structural compatibility with cancer-cell displacement.

**Initial condition**. Tumor seeding was initialized as a compact circular region centered in the tissue field. If $(x_c, y_c)$ denotes the center of the domain, initial malignant occupation was assigned according to

$$M_{ij}(0) = 1 \text{if} (x_i - x_c)^2 + (y_j - y_c)^2 \leq r_0^2 \text{and} P_{ij} = 1,$$

and $M_{ij}(0) = 0$ otherwise. The initial radius was set to approximately $r_0 = 0.16$ mm in the main simulations, corresponding to eight lattice sites at $\Delta x = 20\ \mu$m. This produced a localized malignant seed with a maximum initial geometric area of $\pi r_0^2 \approx 0.080\ \text{mm}^2$, although the realized initial area was smaller when parts of the initial disk overlapped non-permissive tissue. This condition was used to avoid imposing initial global connectivity. Therefore, spatial expansion had to emerge from the interaction between local growth, tissue permissiveness and clearance, rather than being present from the beginning.

**Growth dynamics**. Malignant expansion was updated in discrete daily steps for $T = 21$ simulated days. At each step, candidate sites for colonization were generated by a morphological dilation of the malignant set. If $\mathcal{N}(i, j)$ denotes the von Neumann neighborhood of site $(i, j)$, a site became a candidate if at least one neighboring site was malignant,

$$C_{ij}(t) = 1 \text{if} M_{ij}(t) = 0,\ P_{ij} = 1, \sum_{(k,l)\in\mathcal{N}(i,j)} M_{kl}\,(t) > 0.$$

In simulations including larger motility distances, dilation was repeated $s$times per day, where

$$s = \text{round}\left(\frac{v_m}{\Delta x}\right),$$

and $v_m$ is the motility scale in $\mu$m day$^{-1}$. For example, $v_m = 55\ \mu$m day$^{-1}$ corresponds to approximately three lattice steps per day. Candidate sites were colonized stochastically according to a Bernoulli process,

$$G_{ij}(t) \sim \text{Bernoulli}(\beta C_{ij}(t)),$$

where $\beta$ is the daily colonization probability. The malignant state after growth was

$$M_{ij}^{+}(t) = M_{ij}(t) \vee G_{ij}(t).$$

This rule encoded local expansion into adjacent permissive tissue without assuming long-range jumps or continuous diffusion.

**Clearance dynamics**. Clearance was implemented as a stochastic removal process acting on malignant sites after the growth step. For each malignant site, removal was sampled as

$$R_{ij}(t) \sim \text{Bernoulli}(\gamma M_{ij}^{+}(t)),$$

where $\gamma$ is the daily clearance rate, expressed as the fraction of malignant sites removed per day. The updated malignant state was

$$M_{ij}(t+1) = M_{ij}^{+}(t)\big(1 - R_{ij}(t)\big).$$

The parameter $\gamma$ was varied between $0$ and $0.15$ day$^{-1}$, corresponding to 0 to 15 percent malignant-cell clearance per day. This quantity was not interpreted as a specific therapy or immune mechanism, but as a general local removal term. In some simulations, a low-probability local filling step was included to represent proliferative replacement within permissive neighborhoods,

$$F_{ij}(t) \sim \text{Bernoulli}(\alpha L_{ij}(t)),$$

where $L_{ij}(t) = 1$ for non-malignant permissive sites adjacent to malignant occupation and $\alpha$ was set lower than $\beta$. The final update was constrained again by $P_{ij}$.

**Connectivity measures**. At each simulated day, malignant connectivity was quantified by connected-component labeling on the binary field $M_{ij}(t)$. Two malignant sites were considered connected if they shared a nearest-neighbor edge. The malignant set was decomposed into components

$$\mathcal{M}(t) = \bigcup_{q=1}^{K(t)} \mathcal{C}_q\,(t),$$

where $K(t)$ is the number of disconnected malignant clusters and $\mathcal{C}_q(t)$ is the $q$-th connected component. The area of each component was computed as

$$A_q(t) = |\,\mathcal{C}_q(t)\,|\,\Delta A,$$

where $|\,\mathcal{C}_q(t)\,|$ is the number of occupied lattice sites in that component. Total malignant area was

$$A_{\text{tot}}(t) = \sum_{i,j} M_{ij}\,(t)\Delta A,$$

and the largest connected malignant domain was

$$A_{\max}(t) = \max_q\ A_q(t).$$

Fragmentation was represented by $K(t)$, while spatial dominance was represented by the largest-cluster fraction,

$$\phi_{\max}(t) = \frac{A_{\max}(t)}{A_{\text{tot}}(t)}.$$

These quantities allowed tumor burden and tumor connectivity to be analyzed separately.

**Spanning criterion**. A malignant configuration was classified as percolating when at least one connected malignant component touched two opposite boundaries of the tissue domain. If $\mathcal{L}$, $\mathcal{R}$, $\mathcal{T}$ and $\mathcal{B}$ denote the sets of component labels intersecting the left, right, top and bottom boundaries, respectively, the spanning indicator was defined as

$$S(t) = \begin{cases} 1, & (\mathcal{L} \cap \mathcal{R}) \neq \emptyset \text{ or } (\mathcal{T} \cap \mathcal{B}) \neq \emptyset, \\ 0, & \text{otherwise}. \end{cases}$$

The probability of forming a spanning malignant cluster was estimated by repeated simulations,

$$\hat{P}_{\text{span}}(p, \gamma) = \frac{1}{n} \sum_{r=1}^{n} S_r(T),$$

where $n$ is the number of replicate runs and $S_r(T)$ is the final spanning state in replicate $r$. The approximate critical permissive fraction was estimated as the smallest value of $p$ for which

$$\hat{P}_{\text{span}}(p, \gamma) \geq 0.5.$$

This threshold criterion was used only as an operational numerical definition, not as an analytical proof of a universal critical point.

**Parameter sweeps**. We explored two main control variables: permissive tissue fraction $p$ and daily clearance rate $\gamma$. In the principal sweep, $p$ was sampled from 0.35 to 0.80 and $\gamma$ from 0 to 0.15 $\text{day}^{-1}$. For each parameter pair, multiple stochastic replicates were generated using independent random seeds. For each replicate, the simulation returned $A_{\text{tot}}(T)$, $A_{\max}(T)$, $K(T)$ and $S(T)$. Averaged outputs were then computed as

$$\bar{A}_{\text{tot}}(p, \gamma) = \frac{1}{n} \sum_{r=1}^{n} A_{\text{tot},r}(T),$$

$$\bar{A}_{\max}(p, \gamma) = \frac{1}{n} \sum_{r=1}^{n} A_{\max,r}(T),$$

and

$$\bar{K}(p, \gamma) = \frac{1}{n} \sum_{r=1}^{n} K_r(T).$$

The heatmap of final malignant area displayed $\bar{A}_{\text{tot}}$ in $\text{mm}^2$. The phase diagram displayed $\hat{P}_{\text{span}}$. These quantities were chosen because they distinguish expansion magnitude from spatial continuity.

**Software tools**. Numerical arrays and random sampling were handled with NumPy. Binary dilation and connected-component labeling were performed with SciPy, specifically functions from scipy.ndimage. Data visualization was performed with Matplotlib, including line plots, heatmaps, color bars, image maps and inset axes. The code used pseudo-random number generators initialized with explicit seeds through numpy.random.default_rng.

## RESULTS

We report the outcomes of spatial simulations quantifying tumor expansion, fragmentation and connectivity under varying permissive tissue fractions and clearance rates. Our analysis compared geometric growth metrics with connectivity-based measures to determine whether spatial continuity emerges independently of total malignant burden.

**Spatial regimes**. Simulations revealed three reproducible regimes across parameter space (Figure 1). At low permissive fractions (approximately 35–50%) combined with moderate clearance rates (≥10% per day), malignant growth remained spatially fragmented, with cluster counts remaining high and the largest connected component representing a limited fraction of total malignant area. Under these conditions, total malignant area increased modestly over time but did not produce a spanning structure. At intermediate permissive fractions (approximately 55–65%) and clearance rates between 5% and 10% per day, growth exhibited mixed behavior, with intermittent coalescence of clusters and variable emergence of large connected domains depending on stochastic realization. At higher permissive fractions (≥70%) and low clearance rates (≤5% per day), malignant cells consistently formed a single dominant cluster spanning the tissue domain, with the largest connected component approaching the total malignant area.
Time-course trajectories showed that malignant and largest-cluster areas diverged early in subcritical regimes but converged in supercritical regimes, indicating that connectivity is not trivially determined by tumor size. Cluster counts decreased sharply only in the high-permissiveness regime, reflecting coalescence rather than elimination.
These observations suggest that spatial expansion could depend on the formation of connected malignant paths rather than on absolute cell number alone, establishing a separation between geometric growth and topological continuity.

**Parameter dependence**. Quantitative analysis of the parameter space demonstrated a structured dependence of malignant connectivity on permissive fraction and clearance rate (Figure 2). The heatmap of final malignant area showed a monotonic increase with permissive fraction and a decrease with clearance rate, with area values spanning from approximately 0.2 mm² in low-permissiveness/high-clearance conditions to values exceeding 2 mm² in high-permissiveness/low-clearance conditions. However, comparison with spanning probability revealed that similar malignant areas could correspond to distinct connectivity states. For example, intermediate parameter combinations produced total malignant areas comparable to those observed in high-permissiveness conditions, yet failed to generate spanning clusters. The estimated transition between non-spanning and spanning configurations occurred within a narrow band of permissive fractions, shifting toward higher values as clearance increased. This shift was consistent across replicate simulations, indicating that clearance alters the effective connectivity required for spatial continuity. The largest-cluster fraction increased sharply across this transition, whereas total malignant area changed more gradually, reinforcing that connectivity exhibits threshold-like behavior distinct from overall growth.
Therefore, the emergence of a system-spanning malignant structure is governed by a critical combination of permissive tissue and clearance dynamics, rather than by tumor burden alone.

Overall, tumor spatial growth can be decomposed into two partially independent components: the expansion of malignant area and the emergence of a connected organization. Our simulations indicate that connectivity develops only when permissive tissue and clearance conditions jointly cross a threshold, leading to a transition from fragmented clusters to a spanning configuration. Within this view, containment is associated with the persistence of fragmented spatial organization, rather than with size alone.

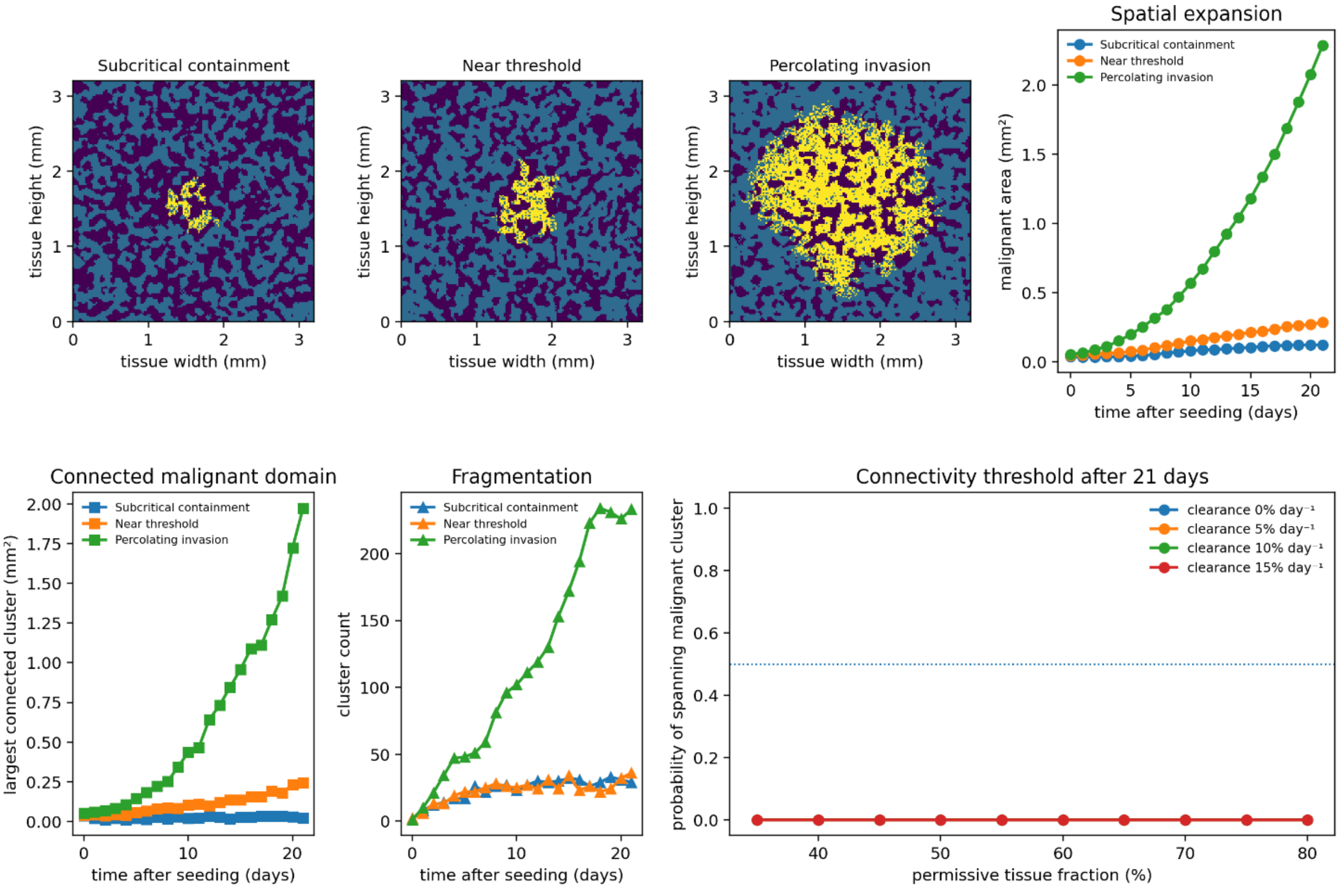

**Figure 1**. Integrated visualization of tumor spatial dynamics in a 3.2 × 3.2 mm tissue domain with spatial resolution of 20 μm per cell over 21 simulated days.

The upper row displays representative spatial configurations under the three regimes of subcritical containment, near-threshold behavior and percolating invasion. In each map, permissive tissue regions are shown as background structure and malignant cells as occupied sites, illustrating how increasing connectivity enables the formation of extended malignant domains. The rightmost panel of the upper row reports total malignant area over time, highlighting differences in expansion kinetics across regimes.
The lower row quantifies structural organization. The left panel shows the area of the largest connected malignant cluster, capturing the emergence of a dominant spanning component; the central panel reports the number of disconnected malignant clusters, reflecting fragmentation versus coalescence; the right panel displays the probability of forming a spanning cluster as a function of permissive tissue fraction and malignant-cell clearance rate, identifying the transition between subcritical and supercritical regimes.
Together, Figure 1 illustrates how spatial connectivity governs the shift from localized growth to system-spanning tumor expansion.

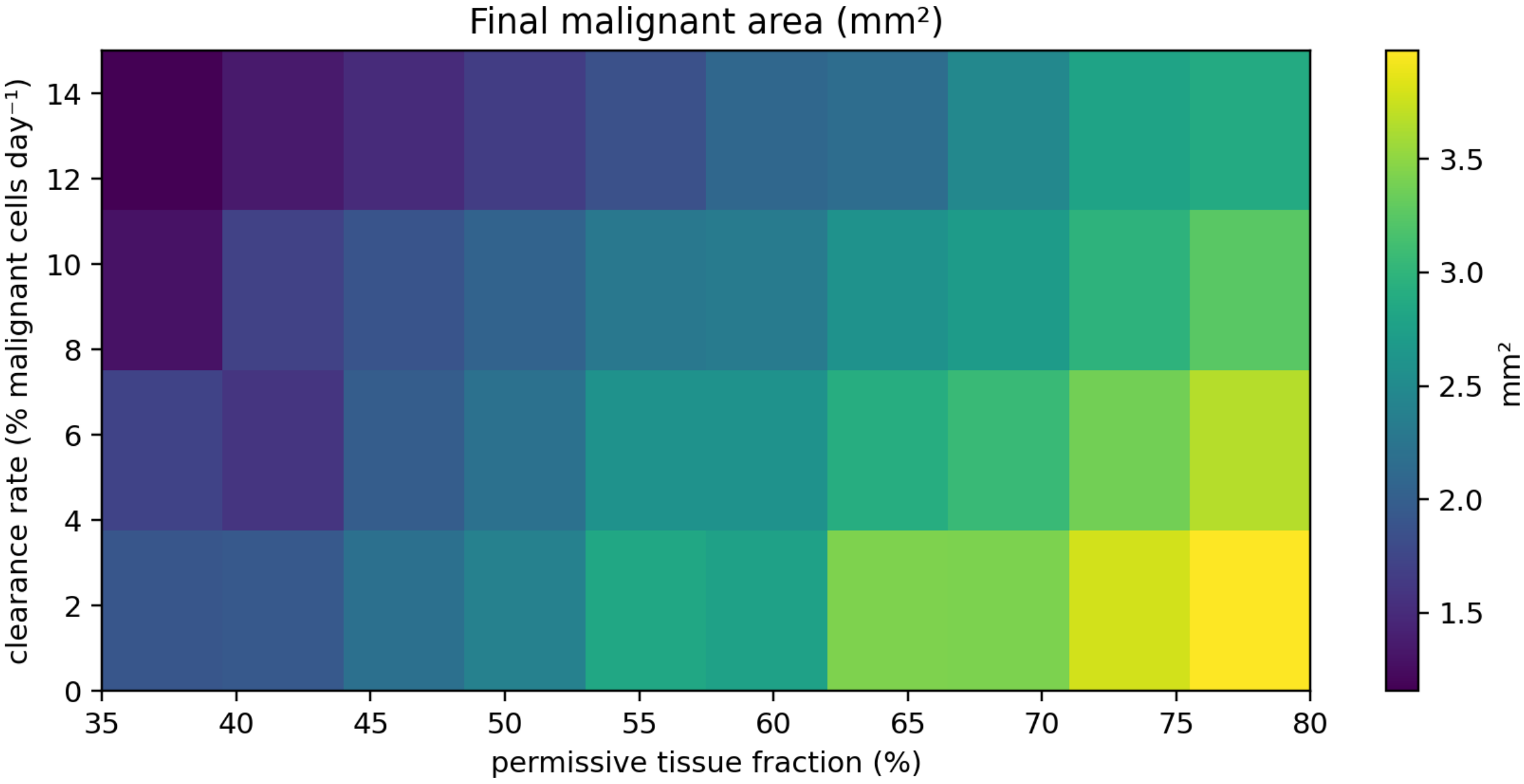

**Figure 2**. Final malignant area as a function of permissive tissue fraction and malignant-cell clearance rate after 21 days of simulated growth in a 3.2 × 3.2 mm tissue domain with spatial resolution of 20 μm per cell. The heatmap reports the total area occupied by malignant cells, showing that increasing tissue permissiveness promotes spatial expansion, whereas higher clearance rates reduce the extent of tumor occupation. The combined parameter space reveals a transition from confined, low-area states to extended, high-area configurations, consistent with connectivity-dependent growth behavior.

## CONCLUSIONS

We asked whether tumor spatial growth can be described in terms of the emergence of a connected malignant domain and whether containment can be interpreted as the absence of this connectivity. Our simulations compared conditions of varying tissue permissiveness and clearance rates within a spatially explicit representation, tracking both total malignant area and the structure of connected components. We found that malignant expansion does not necessarily imply the formation of a spanning cluster and that similar tumor burdens can correspond to markedly different connectivity states. In particular, the transition from fragmented to connected configurations occurred within a restricted parameter region, where the largest connected component rapidly increased relative to total malignant area. This behavior indicates that spatial continuity is governed by a threshold-like condition arising from the interaction between local growth and removal processes. This suggests that tumor progression could depend not only on accumulation, but also on the organization of accessible paths. This perspective provides a way to distinguish between configurations that are similar in size but differ in their structural capacity to sustain large-scale spatial extension.
Our formulation introduces a measure of connectivity that can be computed directly from spatial configurations. This allows the identification of a transitional regime that cannot be captured by aggregate metrics alone and by models based on growth kinetics or diffusion. By linking spatial organization to measurable quantities, we enable comparison between configurations that would otherwise appear equivalent under standard metrics. This situates tumor expansion within a class of problems characterized by percolation and connectivity thresholds from statistical physics (Karrer et al. 2014; Sidorova et al. 2020; Petridou et al. 2021; Deyo 2022; Ishikawa et al. 2023; Wattendorff and Wessel 2024; Yang and Li 2024; Golden and Straley 2024).

Our study has limitations. The stochastic rules governing growth, clearance and motility are phenomenological and not derived from biophysical laws or calibrated experimental datasets. The estimate of a connectivity threshold is based on finite simulations and spanning criteria and is therefore dependent on system size and discretization, without an analytical derivation or explicit uncertainty quantification across lattice scales. The use of Bernoulli processes and discrete dilation provides a tractable approximation of continuous cell dynamics but may distort transport, invasion fronts and interaction ranges. Connectivity is evaluated on a regular lattice with nearest-neighbor relations, which does not fully capture the complexity of real tissue topology or anisotropic microenvironments. The phase diagram relies on smoothed random fields and limited sampling. Open questions include how connectivity measures relate to experimentally accessible imaging data, how thresholds behave under continuous-space formulations and how parameterization could be grounded in quantitative biological measurements.

Several experimentally testable hypotheses can be drawn. First, we hypothesize that the largest connected malignant component, measured from histological sections or 3D imaging reconstructions, exhibits a threshold-like increase as a function of a tissue permissiveness proxy, such as collagen alignment or porosity. Quantitatively, we expect a rapid transition in the ratio $A_{\text{max}}/A_{\text{tot}}$ from values below 0.3 to above 0.7 within a narrow range of this proxy.
Second, we hypothesize that interventions increasing local cell removal, estimated through apoptotic or immune-marker densities, shift the apparent connectivity threshold toward higher permissiveness values; this can be tested by correlating clearance rates (cells·mm $^{-2}$·day $^{-1}$) with the critical fraction of connected regions required for spanning.
Third, we hypothesize that tumor samples with comparable total area $A_{\text{tot}}$ may display distinct connectivity states, such that specimens with $A_{\text{max}}/A_{\text{tot}} < 0.5$ is fragmented, while those exceeding this ratio exhibit continuous domains; this can be evaluated using segmentation-based cluster analysis.
Fourth, we hypothesize that the correlation length of malignant occupation, defined from spatial autocorrelation functions, increases sharply near the transition, providing an independent measurable signature.
Future research may extend these analyses to continuous-space formulations, incorporate anisotropic and heterogeneous tissue architectures derived from imaging data and integrate experimentally measured motility and turnover rates into the model. Additional directions include multi-scale coupling between cellular dynamics and vascular or stromal networks and the development of data-driven parameter inference methods to relate connectivity measures to clinical imaging and histopathological datasets.

Practical implications involve the use of spatial organization metrics to complement existing quantitative assessments. Imaging modalities such as high-resolution histopathology, multiphoton microscopy or contrast-enhanced MRI could be analyzed to extract connected-component statistics, cluster-size distributions and spatial autocorrelation functions. These measurements may support stratification of tissue samples according to structural organization, independent of bulk measures. Computational pipelines integrating segmentation algorithms and graph-based analysis could be implemented to derive connectivity indices from routine diagnostic images. In addition, experimental systems such as organoids or engineered matrices with controlled architecture may be used to monitor how variations in structural properties influence spatial patterns over time.
In conclusion, we show that the tumors' spatial arrangement encodes information beyond aggregate measures. Metrics like the largest connected component and spanning probability enable the discrimination of tumoral tissues with comparable size but different structural organization, distinguishing between fragmented distributions of malignant cells and extended cell assemblies. This approach provides a connectivity-based characterization of tumor architecture, allowing the identification of distinct organizational regimes not captured by conventional size-based metrics.

## DECLARATIONS

**Ethics approval and consent to participate.** This research does not contain any studies with human participants or animals performed by the Author.
**Consent for publication.** The Author transfers all copyright ownership, in the event the work is published. The undersigned author warrants that the article is original, does not infringe on any copyright or other proprietary right of any third part, is not under consideration by another journal and has not been previously published.
**Availability of data and materials.** All data and materials generated or analyzed during this study are included in the manuscript. The Author had full access to all the data in the study and took responsibility for the integrity of the data and the accuracy of the data analysis.
**Disclaimer**. The views expressed are those of the author and do not necessarily reflect those of the affiliated institutions.
**Competing interests.** The Author does not have any known or potential conflict of interest including any financial, personal or other relationships with other people or organizations within three years of beginning the submitted work that could inappropriately influence or be perceived to influence their work.
**Funding.** This research did not receive any specific grant from funding agencies in the public, commercial or not-for-profit sectors.

**Acknowledgements:** none.
**Authors' contributions.** The Author performed: study concept and design, acquisition of data, analysis and interpretation of data, drafting of the manuscript, critical revision of the manuscript for important intellectual content, statistical analysis, obtained funding, administrative, technical and material support, study supervision.
**Declaration of generative AI and AI-assisted technologies in the writing process.** During the preparation of this work, the author used ChatGPT 5.3 to assist with data analysis and manuscript drafting and to improve spelling, grammar and general editing. After using this tool, the author reviewed and edited the content as needed, taking full responsibility for the content of the publication.